# Comparison of electromagnetic field solvers for the 3D analysis of plasmonic nano antennas

Johannes Hoffmann*[a], Christian Hafner[a], Patrick Leidenberger[a], Jan Hesselbarth[a], Sven Burger[b]
[a]Laboratory for Electromagnetic Fields and Microwave Electronics, ETH Zurich, Gloriastrasse 35, 8092 Zurich, Switzerland
[b]Zuse Institute Berlin, Takustrasse 7, 14195 Berlin, Germany

## ABSTRACT

Plasmonic nano antennas are highly attractive at optical frequencies due to their strong resonances – even when their size is smaller than the wavelength – and because of their potential of extreme field enhancement. Such antennas may be applied for sensing of biological nano particles as well as for single molecule detection. Because of considerable material losses and strong dispersion of metals at optical frequencies, the numerical analysis of plasmonic antennas is very demanding. An additional difficulty is caused when very narrow gaps between nano particles are utilized for increasing the field enhancement. In this paper we discuss the main difficulties of time domain solvers, namely FDTD and FVTD and we compare various frequency domain solvers, namely the commercial FEM packages JCMsuite, Comsol, HFSS, and Microwave Studio with the semi-analytic MMP code that may be used as a reference due to its fast convergence and high accuracy.

## 1. INTRODUCTION

Electromagnetic field solvers may be based either on domain discretization or on boundary discretization techniques. The former are more popular and offer both time domain and frequency domain versions. Which of these techniques is most advantageous depends very much on specific properties of the problem to be solved – although software vendors try to offer very general packages, which may handle arbitrary 2D and 3D problems. It will be shown in this paper that the appropriate selection of the field solver may drastically reduce the computation time and memory requirement.

In order to compare various codes, a simple test case consisting of two gold nano spheres (see Fig. 1) is considered in the following. Essentially this setup was experimentally studied in a recent paper [1], where the gap size was modified over a large range. Strong plasmonic interaction is only obtained for very small gaps of a few nanometers. When the gap becomes too small, tunnel currents may destroy the validity of classical Maxwell models with their macroscopic description of materials. It seems that classical Maxwell solvers provide reasonable results for spheres of 20 nm and more in diameter and even for a gap size of only 1 nm. Besides the possible inaccuracy caused by inappropriate macroscopic models, the numerical problems grow considerably when the gap size is reduced. We therefore consider a pair of gold spheres of 80 nm diameter – as in the experiment – and focus on a relatively small gap of 1 nm. In principle, the response of the pair of spheres depends on the incident wave as well as on the observation point, where the response is measured. For biosensing, strong field enhancement in the gap is most attractive. Therefore, we measure the field in the center of the gap and we apply a plane wave excitation that creates maximum electric field in this observation point: The incident plane wave propagates perpendicular to the axis of the pair of spheres and is polarized parallel to the axis. It should be noted that the configuration is axisymmetric, but the excitation breaks this symmetry. The most efficient way to handle this consists in a symmetry decomposition of the incident wave and separate solution of each symmetry component. Essentially, a Fourier series in $\varphi$ direction is performed, where $\varphi$ is the angular direction around the axis of the structure. Most of the commercial codes do not automatically perform such a Fourier decomposition. Since our goal is to compare the 3D performance of codes for the analysis of optical antennas with and without rotational symmetry, we do not take advantage of this symmetry in order to reduce the computation time. There is only one exception: JCMsuite offers automatic Fourier decomposition, which drastically reduces the computation time and allows one to obtain high accuracy. This shows how accurate FEM results can be. Besides rotational symmetry, our test case has two simple symmetry planes that can be handled without symmetry decomposition: the electric field is tangential to the plane spanned by the wave vector of the incident wave and the axis of the structure and it is perpendicular to the plane perpendicular to the axis of the structure in the center of the gap. These two symmetry planes may be modeled as Perfect Magnetic Conductor (PMC) and Perfect Electric Conductor (PEC) by all software packages. As a result, only one quarter of the structure needs to be discretized explicitly.

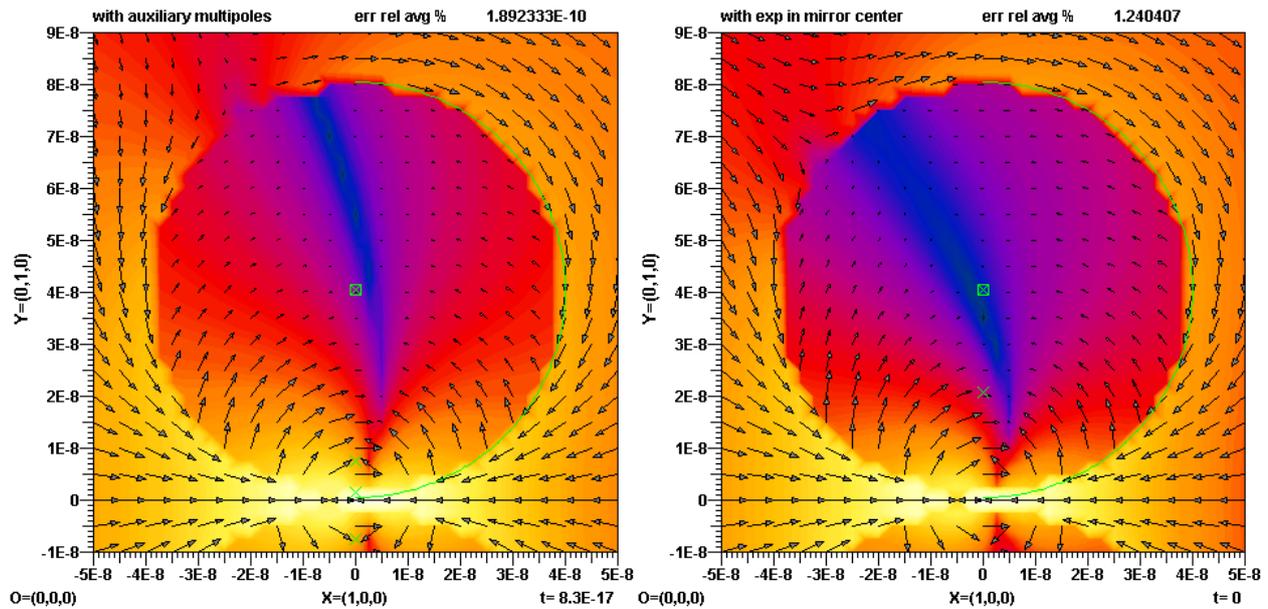

Fig. 1. The time-averaged Poynting vector field at 632 nm wavelength of a plane wave incident on a pair of gold spheres (80 nm diameter each) with a 1 nm gap. The wave is incident in x direction from the left hand side and polarized along the y axis. A logarithmic scale is used for the intensity plot represented by colors because of the huge differences in the field strengths. Left hand side: highly accurate reference solution obtained from a MMP model with 4 auxiliary multipoles (indicated by green x markers) per sphere, maximum degree and order for each expansion 30. Right hand side, quick but not very accurate MMP solution with 2 auxiliary multipoles per sphere, maximum degree and order for each expansion 7 (computation time around 1 s on an AMD Opteron).

There are two features that make the test problem demanding for numerical modeling: 1) The field is strongly enhanced within the tiny gap between the two spheres. This strongly affects interaction of the spheres and the locations of the resonance frequencies of this primitive optical antenna. Finally, this may cause numerical problems. 2) At optical frequencies, gold is lossy and highly dispersive. Consequently, simplified Drude-Lorentz material models [2] are not very accurate and may considerably affect the strengths of the resonance peaks. This mainly affects time-domain solutions, which are always based on Drude-Lorentz approximations or similar material models because these models allow to avoid extremely time-consuming convolution integrals. This difficulty is typical for plasmonics because all of the promising metals are lossy and highly dispersive. However, material dispersion clearly favors frequency domain methods. According to Taflove [3], plasmonic structures require an extremely fine grid of 0.5 nm or less. Since the gap is only 1 nm wide, considerably finer discretization is to be expected. The FDTD code in the MaX-1 software package [4] could not be applied because it is running on personal computers and a personal computer with enough memory was not available. Similarly, it was not possible to run MEEP [5], another well-known FDTD code, on our machines with a sufficiently fine grid due to memory limitations. Other FDTD experts did either not try to solve our test case or they were not able to provide results within months. Roughly one may estimate the FDTD costs by assuming that a cube with a side length of approximately half a wavelength must be discretized to mesh cells of 0.2 nm side length. When no subgridding technique and no local grid refinement are available, one obtains around $2000^3 = 8$ billion grid points for 800 nm wavelength. Depending on the quality and order of the Drude-Lorentz model, one obtains considerably more than 10 real values to be stored per grid point, i.e., the memory requirement is huge. Besides this, very short time steps are required because of the fine discretization, which leads to a large number of FDTD iterations and extremely long computation time.

Since subgridding techniques are often not available or cause additional problems, it is reasonable to concentrate on time domain solvers working on unstructured grids. It is known that standard time domain FEM [6] are very time consuming because a large matrix equation must be solved for each time step. Discontinuous Galerkin (DG) variants of FEM and Finite Volume Time Domain Techniques (FVTD) [7] are much more promising. Unfortunately, we did also not receive DG and FVTD results of our test problem in time. Only frequency domain FEM and MMP results became available within reasonable time. Therefore, this paper mainly contains a comparison of prominent FEM software.

While domain discretization techniques such as FEM lead to relatively large sparse matrices, boundary discretization techniques lead to relatively small but dense matrix equations, which makes them very efficient for 2D simulations. When going from 2D to 3D problems, the discretization domain (the boundaries) becomes 2D rather than 1D, which may drastically increase the matrix size and computation time. As a consequence, boundary discretization techniques are only competitive in 3D modeling when the problem geometry is relatively simple. Since the geometry of our test case is really simple, it may be analyzed by boundary discretization techniques. In fact, the MMP results presented in the following are highly accurate while the corresponding computation time is rather short. Obtaining a reasonable MMP result of a full 3D model takes approximately the same computation time as a Fourier decomposed axisymmetric FEM model. Despite of this, the focus of this paper is on domain discretization techniques, because these techniques are more promising for realistic plasmonic nano antennas with complex geometry. Therefore, we only consider a single boundary discretization technique, the Multiple Multipole Program (MMP) and ignore Boundary Element Methods (BEM) and Boundary Integral Equation Methods (BIEM) [8], which would also be efficient for solving the test case. We show in the following section that MMP provides highly accurate solutions and good error estimates. Therefore, we may use the best MMP solution as reference solution for our FEM comparison.

## 2. MMP BOUNDARY DISCRETIZATION AS REFERENCE SOLUTION

MaX-1 [4] is a software package containing the latest version of the MMP [9] as well as standard FDTD [3] solvers. MMP is a pure boundary discretization technique based on a semi-analytic field expansion, which approximates the electromagnetic field in any homogeneous domain by means of analytic solutions of Maxwell's equations, such as the well known plane waves, multipole and Bessel expansions or more advanced complex origin multipole expansions, ring multipole expansions, line multipole expansions, etc. In the case of Mie scattering, i.e., scattering of a plane wave at a single sphere, MMP may use the same expansions as in the analytic solution, i.e., a multipole expansion for modeling the scattered field and a Bessel expansion for the field inside the sphere. Imposing the continuity conditions for the electromagnetic field on the surface of the spheres then leads to a matrix equation that may efficiently be solved because the Mie expansion is an orthogonal set of basis functions on the surface of the sphere. Since MMP is designed for objects of arbitrary shapes, it uses an advanced error minimization technique that does not assume any orthogonality of the field expansion on the surface. As a result, the MMP solution is more time-consuming but much more general and flexible than the Mie solution. In the case of Mie scattering at a single sphere, however, both methods provide identical results.

For the test problem of a pair of spheres, one may expand the field inside each sphere by a Bessel expansion and the scattered field by two multipole expansions located in the centers of the spheres. When doing this, one must notice that the field of this set of basis functions is no longer orthogonal on the surface of the two spheres. In spite of this accurate results are obtained as long as the interaction between the spheres is weak, i.e., when the gap between the spheres is large enough. Since the gap is chosen to be very small, this Mie type approach becomes inefficient as one can see from Fig. 2. In order to study the convergence, mismatching errors integrated along the boundaries are plotted for increasing orders and degrees of the expansions at 632 nm wavelength, where the maximum field enhancement is observed. Note that this is the most critical wavelength as indicated in a previous 2D study [10]. For wavelengths away from resonance, considerably smaller errors are obtained.

It is interesting to note that exponential convergence should theoretically be obtained. As one can see from Fig. 2, no improvement is obtained when the orders and degrees of the Bessel and multipoles are above 30. The reason for this is that higher order expansions would require higher precision than double precision that is used in the MaX-1 software.

One of the main discoveries of MMP is that the accuracy and efficiency may be further improved by adding additional "auxiliary" multipoles. There is much freedom in the setting of auxiliary multipoles and various automatic routines are available for doing this. Because of the simplicity of the test case, some physical reasoning helps setting the auxiliary multipoles. Obviously, these multipoles should help modeling the strong interaction of the spheres, i.e., mainly the field in the gap area. Therefore, one should add multipole expansions close to the gap on both sides of the gap, once for the interior field and once for the exterior field. Optimal placement of the auxiliary multipoles would be very demanding. It is much easier to set a few multipoles and increase their orders and degrees until the mismatching error along the boundaries of the spheres is reasonable small. As one can see from Fig. 2, already two auxiliary multipoles (one for the interior and one for the exterior field) allow obtaining much better results. The average relative error is now reduced by a factor of 1.8 when the orders and degrees of all expansions are increased by one. Note that simultaneously increasing all orders and degrees is reasonable for a simple convergence study but in general, it is not optimal. An adaptive multipole setting based on an error analysis would allow one to obtain better solutions with less computational costs. Obviously, one observes the same numerical problems for orders and degrees above 30 as for the primitive Mie approach. Thus, the minimum relative error that can be reached with the two auxiliary multipoles is around $10^{-4}$ %. Although this is not bad, the error may be further decreased by either optimizing the locations of the expansions or by adding more auxiliary multipoles. Fig. 2 demonstrates that errors below $10^{-11}$ % may be reached now. Furthermore, the exponential convergence rate is now doubled and an improvement factor of 3.6 per order is obtained.

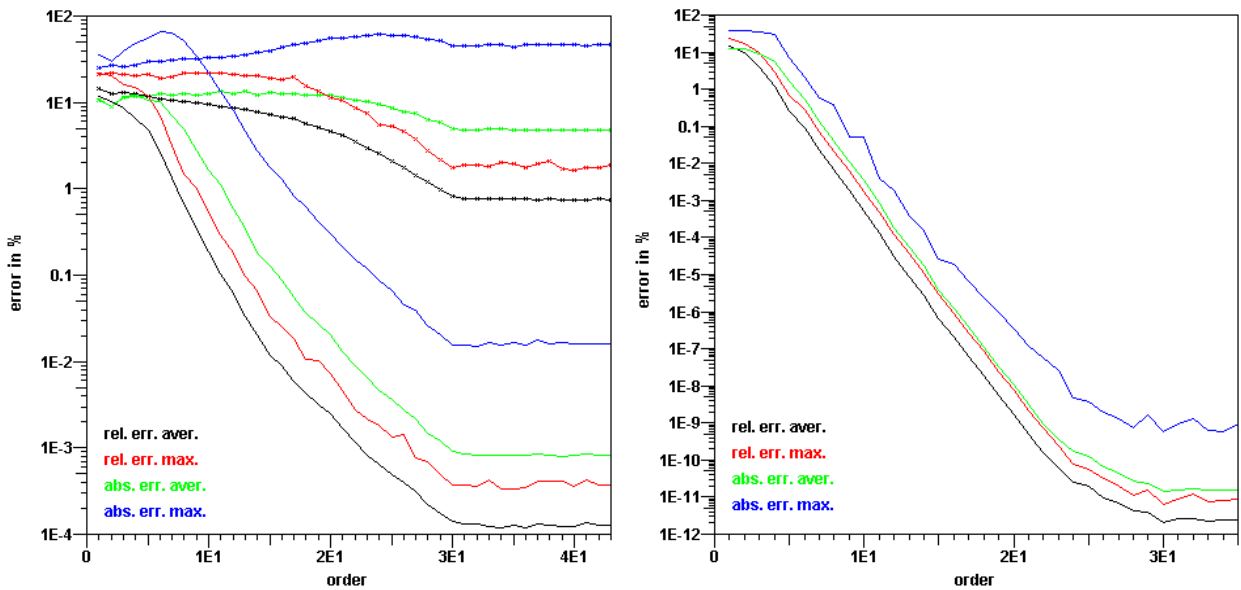

Fig. 2. Convergence of the 3D MMP solver's error. Relative errors are given in % and arbitrary units are used for the absolute errors. Left hand side: Curves with crosses indicate the primitive Mie type approach with Bessel expansion for the interior field and a single multipole expansion for the exterior field; origins of the expansions are in the centers of the spheres. Curves with crosses indicate a slightly improved model with two auxiliary multipoles per sphere. Right hand side: further improvement with four auxiliary multipoles per sphere.

Errors below $10^{-11}$ % are really nice, but it should be mentioned that this is an internal error estimate based on the mismatching errors along the boundaries. These errors contain all six components of the electric and magnetic field, but condensing the information of the error distribution along the boundaries in a single number which is essentially a weighted residual integral hides some details. Therefore, MaX-1 computes both absolute and relative errors and of these the averages as well as the maximum values are plotted in Fig. 2. While relative errors are given in %, arbitrary units are used for the absolute errors. As one can see, the same convergence behavior is observed for all errors, except for the Mie approach, where the convergence region for the maximum absolute error just starts immediately before it is stopped because the maximum order 30 is reached. One may also see that the maximum error is considerably higher than the average error, especially for the absolute errors. This indicates that relatively high errors are obtained near the gap region, where the field is strongly enhanced. Thus, the error of the field in the gap may be larger than the error of the field elsewhere. When the convergence of the field in the center of the gap is considered, one finds that $E^2(0,0)$ converges nicely towards $5.47624*10^5$ $V^2/m^2$ at 632 nm wavelength. It is expected that at least 5 digits of this result are correct. Thus, this MMP computation may be used as reference for FEM and other simulations, which are usually considerably less accurate.

The MMP computation time for order and degree equal to 15 (Mie expansion with four additional multipoles, matrix NxM=1531x14118) was approximately 15 s for the matrix setup and 82 s for the QR solver on a single core AMD Opteron, 64 Bit version. 32 bit versions are usually a bit faster. The setup time is proportional to NxM, the solve time is proportional to $N^2$xM, where N is the number of unknowns and M the number of equations. Heavy overdetermination was used for safety reasons, i.e., M/N is almost 10. With a more careful modeling this factor may be reduced to 2 or even less without reducing the accuracy. This would reduce the computation time by a factor 5. The computation time might be further reduced by a better setting of the multipole locations and by a better selection of their orders.

## 3. COMSOL MULTIPHYSICS

Comsol Multiphysics is a finite element code which allows solving partial differential equations (PDEs) in 2D and 3D domains. There are predefined application modes where the governing PDEs are preset but one can as well freely define the PDEs. Among the predefined application modes there are modules for electromagnetics, mechanics, fluid dynamics, heat transfer and acoustics. All of these application modes can be coupled to multiphysics simulations. Independently of the application mode the geometry, material parameters and boundary conditions of the problem can be set up in a graphical user interface (GUI).

After defining boundaries and domains the mesh has to be generated. This includes the selection of the basis functions' order and the choice of the order of the curved mesh elements. Orders higher than one mean curved elements. They are used for a better approximation of curved boundaries. The Comsol Multiphysics Reference Guide recommends using the same order for the curved elements as for the basis functions. There is an automatic option for the order of the curved elements which is meant to ensure that the order is set to the maximal order of the used basis functions. Nonetheless it was found that this option malfunctions sometimes. Best control over the solution process is achieved by manually setting the order of the curved elements. Another way to influence the meshing is to specify the number of nodes and their distribution on each edge of the model.

As a third mayor step the system of equations has to be solved. Comsol Multiphysics offers a wide variety of solvers for this purpose. There are direct solvers which are suitable for smaller problems and iterative solvers which should be used for larger problems. If there is enough memory available on the computer the direct algorithms can be applied to larger problems as well. This is desirable because they give a more accurate solution than the iterative solvers.

The last task is to visualize or further process the solution. Comsol Multiphysics allows for 2D and 3D plots. Depending on the selected application mode one usually disposes of variables for the electric field and magnetic field. They can be plotted as vector plots and field intensity plots. The from the electromagnetic field derived quantities like Poynting vector, energy and energy loss are as well available as predefined variables. In cases where there is no predefined variable which suits, one can define customized quantities based on the solution vector.

A drawback of Comsol Multiphysics is that it does not allow doing parameter sweeps or optimization within its GUI. For these purposes one has to resort to Matlab. All commands which are accessible from the GUI exist as well as Matlab commands. One can set up a simulation with the GUI and then save it as a Matlab file. The outcome is a script which contains all commands which were issued in the GUI. It is quite easy to change this file in order to parameterize geometries, meshes and every desired aspect of the simulation. Additionally one can use the Matlab optimization toolbox to optimize a given structure.

For the problem at hand a predefined application mode for scattered fields in a 3D domain with harmonic excitation is used. The geometric model, the boundary conditions and material parameters for a wavelength of 632 nm are defined in the GUI. Using the electric and magnetic symmetries allows simulating only one quarter of the whole structure. The bounding box of this quarter has the dimensions 800 nm x 400 nm x 800 nm. The incident wave and the reflected waves are absorbed by a scattering boundary condition. One could as well use perfectly matched layers to obtain better absorption for more oblique incidence, but it is easier to just extend the bounding box a bit and absorb the reflected waves with oblique incidence after multiple reflections. Then the order of the basis functions and of the curved elements is set to two and the problem is meshed with a normal (see Table 1) meshing. As the resulting problem is a quite big matrix, it is solved with the iterative GMRES solver. Next the resulting electric and magnetic fields are visualized and eventually wrong simulation settings can be discovered. To ensure the correctness of the result a convergence analysis is carried out at a wavelength of 632 nm. Table 1 gives the squared electrical field in the center between both spheres in dependence of the mesh density setting, the number of elements and the computation time without meshing.

Table. 1. Convergence study with COMSOL Multiphysics.

| Mesh density | Number of elements / DOF | Computation time [s] | Squared electric field $[V^2/m^2]$ |
|---|---|---|---|
| Extremely coarse | 772/5588 | 5.8 | 703921 |
| Extra coarse | 1637/11966 | 7.9 | 586756 |
| Coarser | 4320/30190 | 16.5 | 586301 |
| Coarse | 6797/46588 | 29.3 | 600013 |
| Normal | 20800/139392 | 79.5 | 576840 |

From this convergence study it can be seen that the result converges and that finer meshes can be quite expensive in terms of computation time. Thus the normal meshing is used for all further computations. As a next step the model is saved as a Matlab file. This Matlab file is extended for the computation of all desired wavelengths. Finally the electric field strength between both spheres can be plotted.

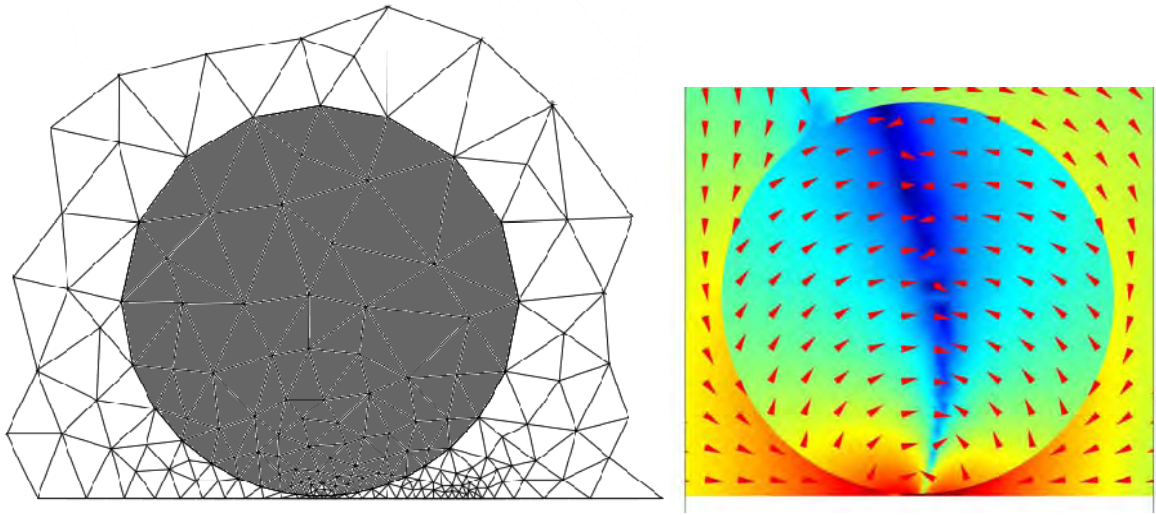

Fig. 3. Left: Magnification of the mesh close to the sphere. The mesh around the sphere appears to be non-conform but the fact that curved finite elements are used compensates for the coarse meshing. Right: Time average of the pointing vector at 632 nm wavelength. Blue indicates small power flows whereas red indicates large power flows. The color scheme is logarithmic. The vectors indicate the direction of the power flow. They are of identical size.

## 4. JCMSUITE

JCMsuite is a software package containing dedicated finite-element (FEM) solvers for simulation tasks appearing in nanooptics [11]. JCMsuite incorporates light scattering solvers, propagation mode solvers and resonance mode solvers. The time-harmonic scattering, eigenmode and resonance problems can be formulated on 1D, 2D and 3D computational domains. Admissible geometries can consist of periodic or isolated patterns, or a mixture of both. Further on, solvers for problems posed on cylindrically symmetric geometries are implemented. The electromagnetic fields are discretized with higher order edge elements. The programme package contains an automatic mesh generator, goal-oriented error estimators for adaptive grid refinement, domain-decomposition techniques and fast solvers. JCMsuite has been applied to a wide range of electromagnetic field computations including metamaterials [12], photonic crystal fibers [13], nearfield-microscopy [14], and optical microlithography [15].

Figure 4 shows a visualization of the Poynting vector field computed from the electric field solution at 632 nm wavelength. Table 2 shows the convergence of the intensity at the origin at 632nm wavelength, computed with JCMsuite towards the quasi-exact value computed with MMP. An accuracy of about one percent is reached within a few seconds computation time and an accuracy of about 0.1% is reached within few minutes of computation time on a standard PC.

For the data displayed in Figure 7, the typical computation time for a single frequency point of the frequency scan is 30 seconds (single-processor usage, standard PC), memory usage is about 0.25 GB of RAM. During the computation the coarse mesh as shown in Fig. 4 is adaptively refined in two successive iterations. Finite elements of third polynomial order are used for the discretization of the electric field.

Table. 2. Convergence study with JCMsuite.

| Program  | Squared electric field $[V^2/m^2]$ | DOF    | Computation time [s] |
|----------|-----------------------------------|--------|----------------------|
| JCMsuite | 5.447417e+05                      | 15552  | 7                    |
| JCMsuite | 5.471188e+05                      | 57828  | 28                   |
| JCMsuite | 5.473468e+05                      | 216572 | 114                  |
| Max-1    | 5.476240e+05                      | 1531   | 97                   |

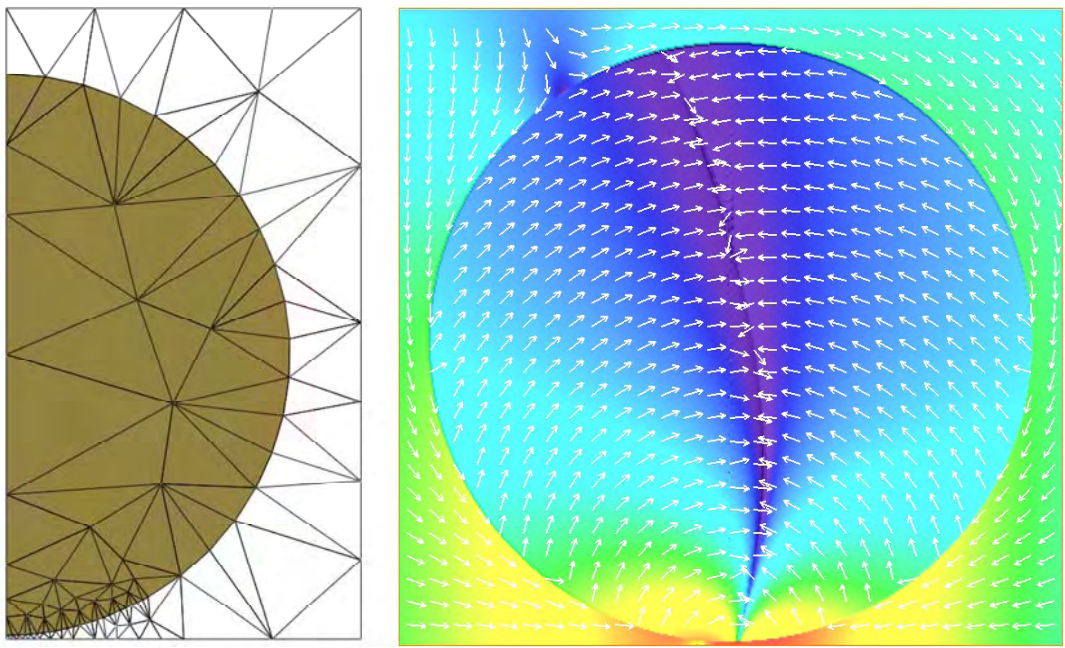

Fig. 4. Left: Coarse mesh for the FEM simulation with JCMsuite. The coarse mesh of the computational domain contains 175 triangles. It is refined adaptively during the field computation to a mesh with about 750 triangles. The left boundary is the rotation axis of the cylindrically symmetric setup, at the lower boundary a boundary condition reflecting mirror symmetry is applied, and at the right and upper boundaries, transparent boundary conditions are applied. Right: A visualization of the Poynting vector field computed from the electric field solution at 632 nm. The intensity plot is in a logarithmic scale.

# 5. HFSS

HFSS is a 3D full-wave finite element (FE) electromagnetic field simulator for passive structures operating in the frequency domain from Ansoft. HFSS is originally dedicated for simulations of RF, microwave and millimeter-wave devices as e.g. PCB interconnects, antennas or microwave components. It has a comfortable GUI for creating the simulation structure, setting up solution setups and parameter studies and allows for solution visualization.

For a new simulation project one needs to define the geometry of the structure first. The geometry is either modeled in the HFSS GUI or can be imported from another program (AutoCAD, STEP and many other file formats are supported). Then one has to assign material properties, boundary conditions and excitations to the different domains and surfaces of the geometry. If desired one can set parameters for the meshing process. One can restrict the length and number of the finite elements for a selected region or surface. Further more it is possible to specify the accuracy with which a defined geometry part should be resolved during meshing process. This is an important feature for curved objects.

To run the simulation one has to specify the solution parameters. The most important parameters are the solution frequency, the order of the base functions and the solver type. Per default HFSS uses a direct solver which is quite fast and memory efficient, but an iterative solver can be used as well. The solution process itself is adaptive. First a solution for an initial mesh is calculated and the solution is rated. If the rating is not good, the mesh will be refined and a new solution will be computed. This procedure will be repeated until one of the exit criteria is achieved. Finally the solution, i.e., the electric and magnetic fields, currents or S-parameters can be visualized in 1D, 2D and 3D. All solutions can be exported as files.

HFSS offers the 'Optimetrics' toolbox for optimizations, parameter, sensitivity and statistical studies. For a more advanced use, HFSS can be programmed with 'Visual Basic Script' and 'Java Script'. HFSS offers no curved elements for a better approximation of curved objects. So if a precise representation of an object is needed, the mesh along the curved surface has to be very fine. This can be achieved with the 'surface approximation' parameter in the mesh operations.

The test problem is modeled using the electric and magnetic symmetry planes. On the remaining surfaces the radiation boundary is applied. Simultaneously this boundary type allows specifying the incoming plane wave with a magnitude of 1 V/m. The computational domain has the size 400 nm x 200 nm x 200 nm. The frequency dependent material parameters for gold can be directly imported to HFSS and associated with the sphere. The direct matrix solver and basis functions of the 2nd order are used to solve the problem.

For the given problem the resolution of the sphere's geometry is very essential for the accuracy of the result. One has to set manually the 'surface approximation' value for the sphere's surface in the mesh operation menu. The influence of the surface approximation value to the solution is shown in Fig. 5 and Table 3. For an inaccurate surface approximation the result does not match with the reference from the MaX-1 code. If one increases the surface resolution using more finite elements at the surface of the sphere, the result converges against the reference. A surface approximation of 0.02 nm which provides an accurate result in acceptable computation time is used for the comparison with other codes.

Table. 3. Convergence study with HFFS.

| Surface resolution [nm] | Number of elements box | Number of elements sphere | Number of elements total | memory consumption of the solver [MB] | time for matrix solution [s] | $\|E\|^2$ [$V^2/m^2$] |
|---|---|---|---|---|---|---|
| 1.0 | 1691 | 421 | 2112 | 179 | 8 | 132842 |
| 0.5 | 1878 | 456 | 2334 | 206 | 11 | 147412 |
| 0.1 | 5026 | 2395 | 7421 | 659 | 55 | 387689 |
| 0.05 | 6226 | 3450 | 9676 | 799 | 65 | 501821 |
| 0.02 | 17290 | 9560 | 26850 | 2370 | 312 | 546820 |
| 0.01 | 26434 | 17526 | 43960 | 3640 | 504 | 537089 |
| 0.008 | 32260 | 22146 | 54406 | 4370 | 629 | 556290 |

In Table 3 the number of finite elements used, the computation time and the memory consumption for a wavelength of 632 nm are listed. The computation time is the pure CPU time for solving the finite element matrix equation with the direct solver. All simulations have been carried out on a PC with an dual core AMD Opteron 2220 with 2.8 GHz and 8 GB memory.

Figure 5 shows the time average of the pointing vector for a wavelength of 632 nm and a surface resolution of 0.02 nm. The mesh used for this simulation is drawn in black. A lot of finite elements are used along the surface of the sphere to approximate the shape well. This is necessary due to the fact that HFSS does not support curved elements. Thus further improvements of the mesh are possible: The essential part of this problem is near the origin of the coordinate system. A very fine surface approximation in this region and a coarser surface approximation on the other parts of the sphere should reduce the number of elements without loss of accuracy.

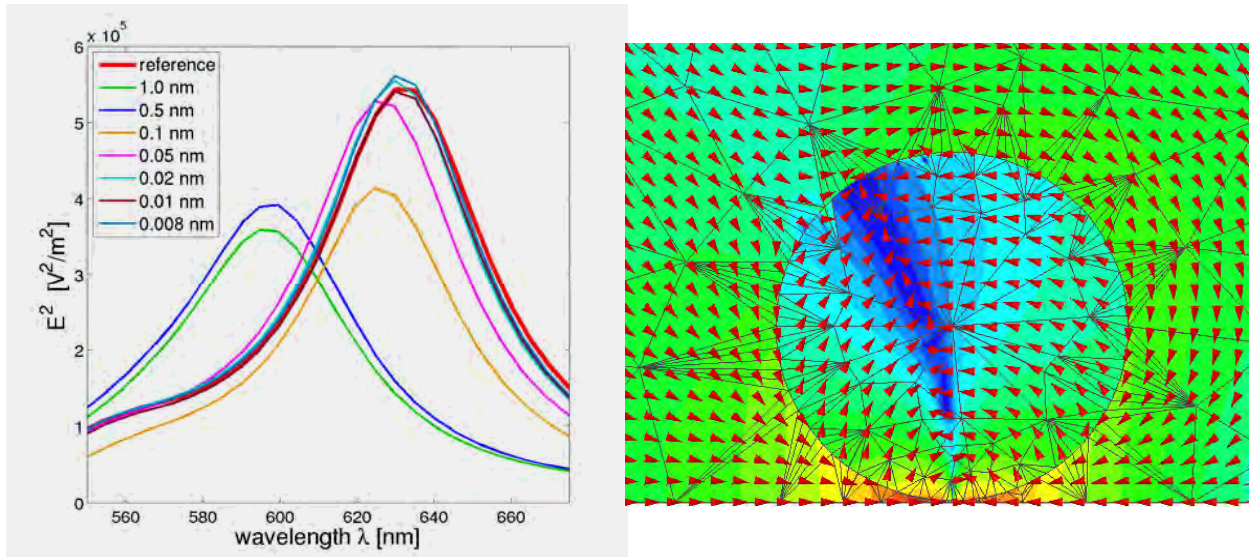

Fig. 5. On the left the influence of the polygonal approximation of the sphere on the simulation result is studied. The given distances are the maximum distances between mesh and sphere. For a coarse resolution of the spherical shape the result differs a lot from the reference, generated with MaX-1 (red line). With a better resolution the solution converges to the reference. On the right the time average of the pointing vector at a wavelength of 632 nm is plotted with a logarithmic scale. The color indicates the magnitude of the pointing vector and the red arrows the direction of the power flow. The mesh was generated for a surface approximation of the sphere of 0.02 nm.

## 6. CST MICROWAVE STUDIO

The test problem is analyzed using the finite-element frequency-domain solver of CST Microwave Studio (version 2008.06). The structure is modeled using two 48-segment spheres. Manual mesh definitions are applied to generate the fine mesh in the vicinity of the gap between the spheres. Figure 6 shows the mesh plot in the symmetry plane. Using both electrical and magnetic symmetry of the structure, the computational domain has been reduced to a quarter of the original problem. On the remaining boundaries an open boundary of type „4 Layer Convolution PML" with the „add space" option is applied. About 100'000 tetrahedra are used to model the problem. Using the iterative solver, calculations for one frequency point took about 5 minutes (PC with AMD Opteron 2.2 GHz) with roughly 2 GB of memory usage.

A number of problems has been encountered while simulating this structure:

- The automatic mesh refinement is useless. Typically, less than 10 tetrahedra are added (or subtracted) during one iteration.
- Spheres with more than 48 segments (which would result in a smoother geometry in the gap region) consistently lead to program crashes.

- Field probes (such as for the E-field in the coordinate origin) do not exist. The field strength must be figured out from 2D field plots.

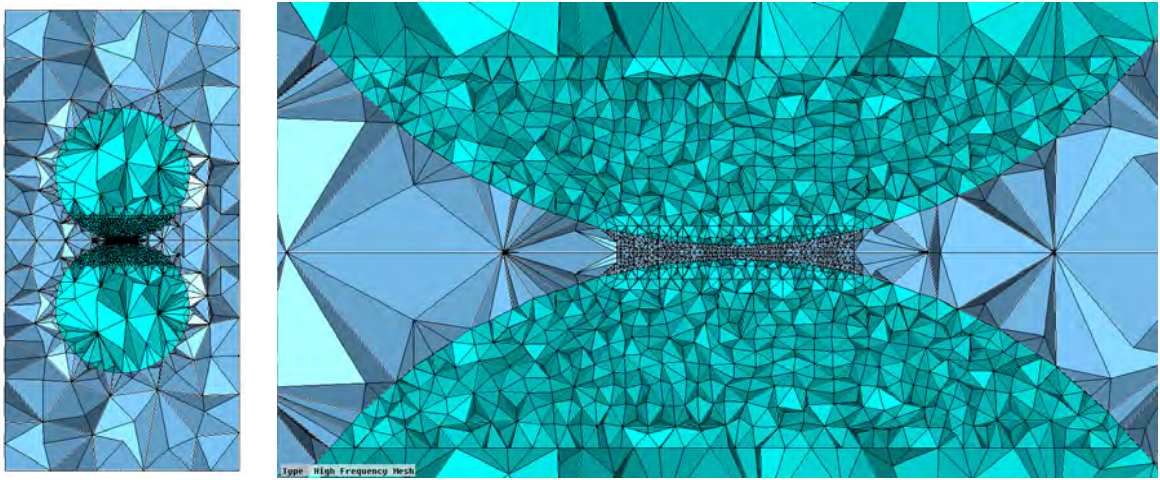

Fig. 6. Tetrahedral FEM mesh plot in the symmetry plane, as generated by CST microwave studio.

## 7. CONCLUSION

The test problem was solved with 4 different FEM codes and the semi-analytic MMP contained in the MaX-1 package. Results from the MMP can be taken as a reference because the boundary error of the best MMP solution can be quantified to $10^{-11}$ percent. Figure 7 shows that essentially results of three different degrees of accuracy were obtained.

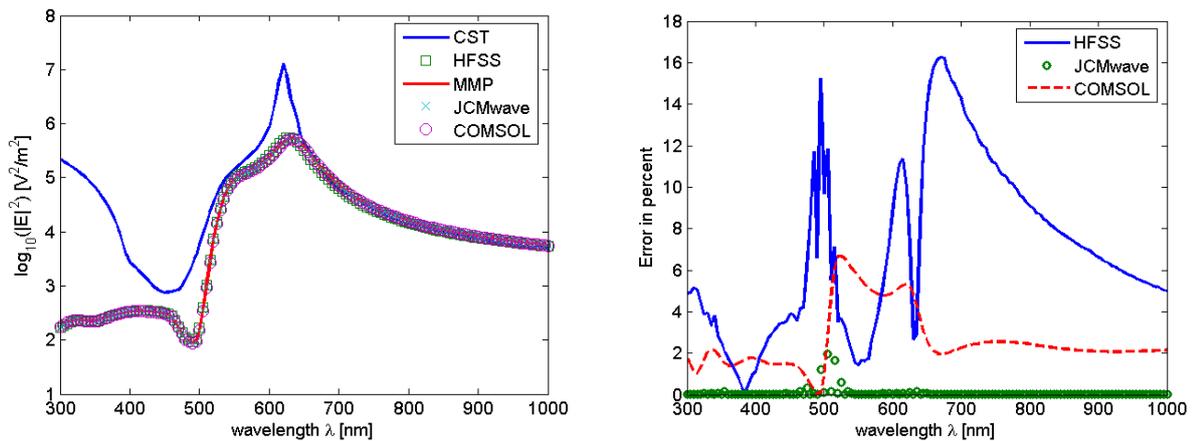

Fig. 7. On the left the resulting field strength in the center between the spheres of all tested codes is compared. All results are quite similar on this level of detail except the CST Microwave Studio result. The right hand side figure shows the relative error of the more accurate solutions. Maximum errors can be observed in regions with steep frequency response. Relative errors larger than 0.1 % in the JCMsuite results are due to a slight difference in the interpolation of the material parameters.

CST Microwave Studio's finite element solver produces obviously inaccurate results. The main reason seems to be the insufficient resolution of the sphere into only 48 segments. More segments consistently produced program crashes. This poor resolution of the sphere's boundary results in a geometry peak in the region where maximum field is expected. The theoretically resulting field singularity falsifies the results.

A much better degree of accuracy is provided by Comsol Multiphysics and HFFS from Ansoft. Again results from both solvers depend on the accuracy of the sphere's polyhedral representation. A secondary influence is the size of the surrounding air box. This size has a slight impact on the resonance frequency. A compromise between size of the box and the required number of elements has to be found.

Best accuracy can be obtained with the JCMwave package. It must be mentioned that the results presented here rely on a Fourier decomposition of the axi-symmetric geometry. The resulting 2D problems can be solved much more accurately than the full 3D problems within much shorter time. The accuracy of the spheres polyhedral representation is not a big problem because of the used Fourier decomposition.

When strong field localization is present – as in the gap between two plasmonic particles – unstructured grids are highly important. With FDTD codes based on structured grids, we were not able to obtain acceptable results within reasonable time. Higher order curved elements are desirable for the simulation of optical antennas and plasmonic structures in general.

The strong material dispersion of metals at optical frequencies clearly favors frequency domain codes. Smart material models may improve the quality of the results of time domain codes, but these models increase the complexity and memory requirement of the codes, while speeding up frequency domain codes by means of Model Based Parameter Estimation (MBPE) [16] and similar techniques is rather simple.

## Correction Note

After publication it was found that wrong material settings for the CST MICROWAVE STUDIO (CST MWS) simulations were used, which lead to erroneous results. Because of the material dispersion, manual frequency sweep was used. A complex $\varepsilon = \varepsilon' + i\varepsilon''$ value was taken from a table that was used for all codes. CST offers input of complex $\varepsilon$ in a table or $\varepsilon'$ and $\tan(\delta)$ instead of complex $\varepsilon$. The second option, which was used in our paper, is only intended for normal dielectric material with *$\varepsilon' \geq 1$*. For negative $\varepsilon'$ a positive $\tan(\delta)$ – CST MWS does not accept negative $\tan(\delta)$ - leads to an "active" material, which explains the much too strong frequency response in the presented CST MWS results. For obtaining correct CST results, one must take advantage of defining complex $\varepsilon$ in a table.

In the paper the spheres were segmented before meshing. This is neither recommended nor required. In order to get an accurate solution using CST MWS 2008, the mesh should be refined before the simulation. For this purpose, global and local mesh parameters should be used. Since the FD Solver uses a tetrahedral mesh, round geometry is segmented during meshing, which creates a local geometry error. In CST MWS 2008 the applied mesh adaption is based on the initial discretization mesh and did not refine the geometry during mesh adaption. Starting with CST MWS 2009, a true geometry adaptation has been added to the tetrahedral frequency domain solver that improves geometry adaptation during mesh adaptation.

After correcting the model and using the correct material settings, the CST MWS 2008 simulation results are close to the reference solution as one may see from the figure below.

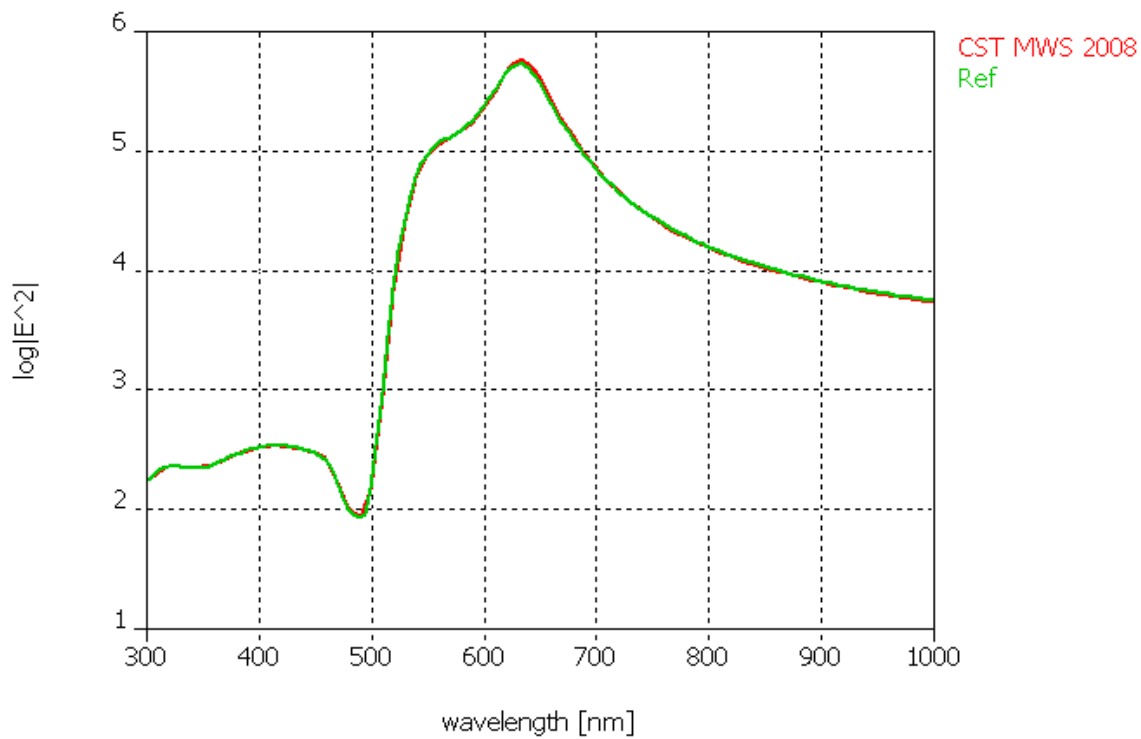

It was mentioned in the paper that field probes were not available. However, the result could be extracted automatically using template based post processing. The "evaluate on point" option works similarly to a probe.

The authors would like to thank CST support for the valuable hints and corrections.